\documentclass[twocolumn,english,prl]{revtex4}
\usepackage[T1]{fontenc}
\usepackage[latin9]{inputenc}
\usepackage{color}
\usepackage{amsmath}
\usepackage{graphicx}

\makeatletter
\@ifundefined{definecolor}
 {\usepackage{color}}{}
\@ifundefined{definecolor}
 {\@ifundefined{definecolor}
 {\usepackage{color}}{}
}{}
\@ifundefined{definecolor}
 {\@ifundefined{definecolor}
 {\@ifundefined{definecolor}
 {\usepackage{color}}{}
}{}
}{}
\@ifundefined{definecolor}
 {\@ifundefined{definecolor}
 {\@ifundefined{definecolor}
 {\@ifundefined{definecolor}
 {\usepackage{color}}{}
}{}
}{}
}{}
\@ifundefined{definecolor}
 {\@ifundefined{definecolor}
 {\@ifundefined{definecolor}
 {\@ifundefined{definecolor}
 {\@ifundefined{definecolor}
 {\usepackage{color}}{}
}{}
}{}
}{}
}{}
\@ifundefined{definecolor}
 {\@ifundefined{definecolor}
 {\@ifundefined{definecolor}
 {\@ifundefined{definecolor}
 {\@ifundefined{definecolor}
 {\@ifundefined{definecolor}
 {\usepackage{color}}{}
}{}
}{}
}{}
}{}
}{}
\@ifundefined{definecolor}
 {\@ifundefined{definecolor}
 {\@ifundefined{definecolor}
 {\@ifundefined{definecolor}
 {\@ifundefined{definecolor}
 {\@ifundefined{definecolor}
 {\@ifundefined{definecolor}
 {\usepackage{color}}{}
}{}
}{}
}{}
}{}
}{}
}{}
\@ifundefined{definecolor}
 {\@ifundefined{definecolor}
 {\@ifundefined{definecolor}
 {\@ifundefined{definecolor}
 {\@ifundefined{definecolor}
 {\@ifundefined{definecolor}
 {\@ifundefined{definecolor}
 {\@ifundefined{definecolor}
 {\usepackage{color}}{}
}{}
}{}
}{}
}{}
}{}
}{}
}{}

\makeatother

\usepackage{babel}

\makeatother

\usepackage{babel}

\makeatother

\usepackage{babel}

\makeatother

\usepackage{babel}

\makeatother

\usepackage{babel}

\makeatother

\usepackage{babel}

\makeatother

\usepackage{babel}

\makeatother

\usepackage{babel}

\begin{document}

\title{Quantum Walk of Two Interacting Bosons }

\author{Yoav Lahini, Mor Verbin, Sebastian D. Huber, Yaron Bromberg, Rami
Pugatch and Yaron Silberberg}

\affiliation{$^{1}$Department of Physics, Weizmann Institute of Science, Rehovot,
Israel.}
\begin{abstract}
We study the effect of interactions on the bosonic two-particle quantum
walk and its corresponding spatial correlations. The combined effect
of interactions and Hanbury-Brown Twiss interference results in unique
spatial correlations which depend on the strength of the interaction,
but not on its sign. The results are explained in light of the two-particle
spectrum and the physics of attractively and repulsively bound pairs.
We experimentally measure the weak interaction limit of these effects
in nonlinear photonic lattices. Finally, we discuss an experimental
approach to observe the strong interaction limit using single atoms
in optical lattices.
\end{abstract}
\maketitle
\emph{Introduction}\textbf{\emph{.-}} Understanding highly correlated
many body systems remains both an experimental and theoretical challenge.
While there is a rather good understanding of weakly interacting systems,
problems involving strong interactions are in general harder to address.

Recently a new approach to the study of quantum dynamics became experimentally
accessible through the study of Quantum Walks (QWs) in lattice potentials
\cite{Qwalk}. Quantum walks are the quantum counterparts of classical
random walks on discrete lattice: A quantum particle is initially
placed at a particular site of a lattice and then tunnels to neighboring
sites with equal probability amplitude. This basic {}``step'' is
repeated, but in contrast to the classical case quantum mechanical
interference leads to distinctively different dynamics. For example,
in periodic lattices the wavefunction width grows ballistically, while
in the classical case the expansion is diffusive.

QWs receive increasing attention due to their relation to various
physical and bio-physical processes \cite{Qwalk,Photosynthesis,Takuya},
and their possible use as a primitive for quantum computation algorithms
\cite{UniversalQC}. Theoretically, QWs were studied for the single
particle case \cite{Qwalk}. Initial experiments studied the physics
of single particles by using either classical waves \cite{hagai},
single photons \cite{Silberhorn,White}, or single atoms \cite{Bloch,atomQwalk}.
Moving from one to two non-interacting particles it has recently been
shown that indistinguishable quantum walkers can develop non-trivial
correlations due to Hanbury Brown-Twiss (HBT) interferences \cite{Bromberg,Peruzzo,HBT}.
Yet, very little is known on the effect of interactions on the dynamics
of the few-body QW \cite{Havlin}. As new systems emerge that can
accommodate such experiments \cite{Bloch,atomQwalk}, a systematic
study of this problem starting at small particle numbers may offer
a {}``bottom up'' approach in the general thrive to understand dynamical
quantum many-body systems.

In this letter we study the effect of inter-particle interactions
on the two-particle quantum walk and the resulting spatial correlation.
We consider two bosons, each initially localized on a single lattice
site, undergoing a QW simultaneously. We find that the interplay between
interactions and quantum two-particle (HBT) interference gives rise
to fermion-like spatial correlations between the particles. Interestingly,
the correlations depend on the strength of the interaction but not
on whether it is attractive or repulsive. We explain the observed
correlations by calculating the two-particle spectrum, and interpret
our results in light of the physics of attractively and repulsively
bound pairs \cite{Winkler,Maldague}. We then present an experimental
observation of the weak interaction limit of these effects in nonlinear
photonic lattices, and outline an experimental approach to observe
the strong interaction limit using single atoms in optical lattices.

\begin{figure}
\includegraphics[clip,width=3.4in]{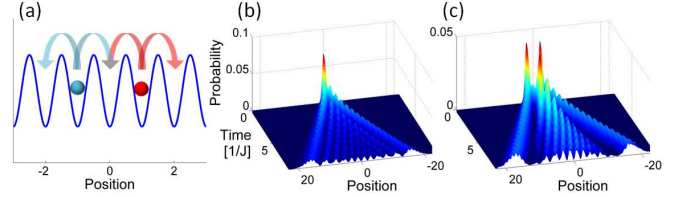}

\caption{\label{fig: 1-1}(color online). Two-particle Quantum Walk. (a) An
illustration of two identical atoms initially placed on two sites
of an optical lattice, and allowed to tunnel to neighboring sites
and interfere. (b) The evolution of the particle-density for the case
of a single atom quantum walk. (c) The evolution of the density for
a two-atom quantum walk. This density is only weakly affected by interactions,
while the two particle correlations are strongly modified.}

\end{figure}

\begin{figure}
\includegraphics[clip,scale=0.8]{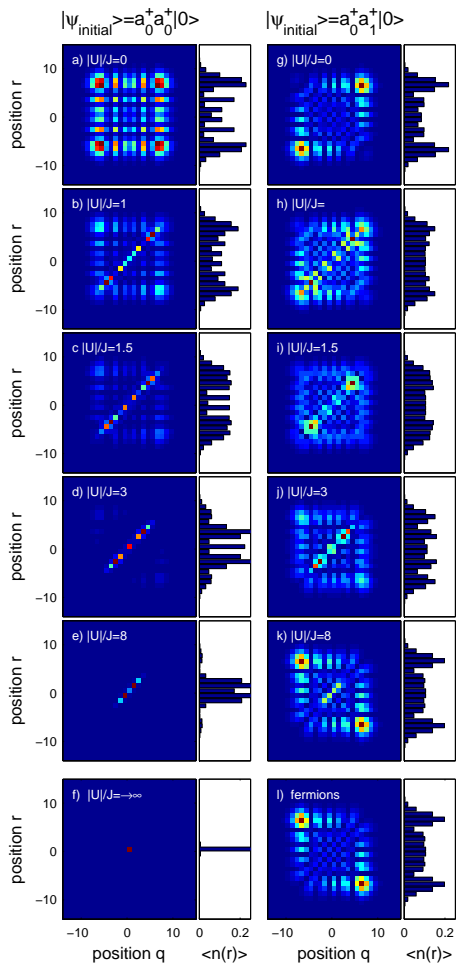}

\caption{\label{fig: 1}(color online). Two-particle correlations for
interacting quantum walkers. Left column: correlations after
propagation time $T=4$, where initially the two particles are placed
at the same site,
$|\psi_{initial}\rangle=(a_{0}^{\dagger})^{2}|0\rangle$. (a) For
zero interactions, the two-particle correlation shows no
interference \cite{Bromberg}. (b)-(f) As the interaction is
increased the correlations shows the formation of bound pairs, while
the density distribution (shown on the right of each plot) becomes
increasingly localized. Right column: similar results for an initial
condition in which the two particles are placed at adjacent sites
$|\psi_{initial}\rangle=a_{1}^{\dagger}a_{0}^{\dagger}|0\rangle$.
(g) At $U=0$ the correlations shows spatial bunching. (h)-(j) The
correlations change as the interaction $|U|$ is increased, while the
density is only weakly affected. (k) At strong interactions the
correlation is transformed to spatial anti-bunching, similar to the
correlation that would be exhibited by two non-interacting fermions
initially placed in the same configuration (l). All results are
identical for both attractive and repulsive interactions.}

\end{figure}

\emph{QW of two interacting particles}.- We start by calculating the
QW of two interacting particles. We consider the one dimensional Bose-Hubbard
model:

\begin{equation}
H=-J\sum_{\langle l,m\rangle}a_{l}^{\dagger}a_{m}+\frac{U}{2}\sum_{m}\hat{n}_{m}(\hat{n}_{m}-1)\label{eq:QH}\end{equation}
 where $a_{m}^{\dagger}$ $(a_{m})$ is the creation (annihilation)
operator for a particle at site $m$, $\hat{n}_{m}=a_{m}^{\dagger}a_{m}$
is the corresponding number operator, $J$ is the tunneling amplitude
between nearest neighbors and $U$ is the on-site interaction energy
which can be attractive (negative) or repulsive (positive).

We study the QW of two indistinguishable particles, each initially
localized on a single site in a periodic lattice. We consider two
different initial conditions: One in which the two particles are localized
at adjacent lattice sites $|\psi_{initial}\rangle=a_{1}^{\dagger}a_{0}^{\dagger}|0\rangle$,
and a second in which the particles are initially placed at the same
site, $|\psi_{initial}\rangle=(a_{0}^{\dagger})^{2}|0\rangle$. Our
focus lies on the particle-density $n_{r}(t)$=$\left\langle a_{r}^{\dagger}a_{r}\right\rangle $
and on the two-particle correlation $\Gamma_{q,r}(t)=\left\langle a_{q}^{\dagger}a_{r}^{\dagger}a_{r}a_{q}\right\rangle $
which are calculated after an evolution time $T$ for different values
of the interaction $U$. At all stages the particles are far from
the lattice boundaries.

The results for $U=0$ correspond to the results reported in \cite{Bromberg,Peruzzo}:
when the two bosons start the QW at the same site, after propagation
each particle can be found on either side of the site of origin, reflected
in the four symmetric peaks in the correlation matrix inset for Fig.
2a.When the particles are initially placed at adjacent sites, HBT
interference results in spatial bunching, and the two particles propagate
together either to the left side of the distribution (peak at the
bottom left corner of the correlation matrix in \ref{fig: 1}g) or
to the right (top right corner). Other initial conditions, in which
the particles are further separated in space result in more complicated
correlation patterns \cite{Bromberg}. Such initial state and the
results of interactions will be discussed in \cite{Sup}.

Let us now turn to the discussion of interaction effects. Fig. \ref{fig: 1}
b-e show the results for increasing repulsive interaction $U$ for
in the case of two bosons initially localized at the same site. The
spatial correlations show that as $|U|$ increases the two particles
tend to propagate as a pair, while the density distribution becomes
localized. Fig. \ref{fig: 1} g-l shows the results for the case in
which the particles are initially places at different sites$|\psi_{initial}\rangle=a_{1}^{\dagger}a_{0}^{\dagger}|0\rangle$.
Here, the particle density depends only weakly on $U$, but the two-particle
correlation undergoes a fundamental change: the spatial bunching effects
which occur in the non-interacting case gradually transforms to spatial
anti-bunching (Fig. \ref{fig: 1} k). For large values of the interaction
strength $|U|$, the correlation between the two bosons becomes very
similar to the correlation exhibited by two non-interacting fermions,
prepared in the same initial configuration (compare Fig. \ref{fig: 1}
k and l). The non-interacting fermionic and the interacting bosonic
matrices become identical at the limit of $|U|\rightarrow\infty$,
while the density becomes identical to the one for $U=0$. An interesting
result is that in both cases the effect of interactions does not depend
on the sign of $U$; it is identical for both attractive and repulsive
interactions. We note that for initial conditions in which the two
particles are further separated in space interactions also drive the
system towards fermion-like correlations, only that now they have
a more complicated spatial structure- see \cite{Sup} for additional
experimental and theoretical results.

\begin{figure}
\includegraphics[clip,scale=0.8]{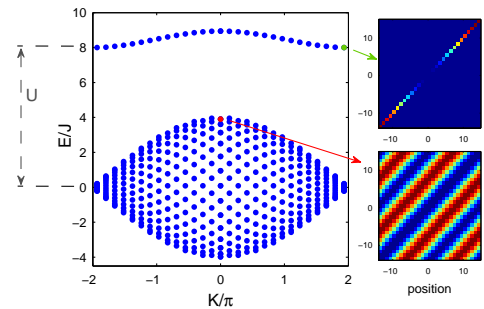}

\caption{\label{fig: 2}(color online). Spectrum and two-particle eigenmodes
of Eq. \ref{eq:CH} for $U=8.$ The spectrum is separated into two
mini-bands. The higher band consists of bound pair states in which
the two particles only occupy the same sites (non-zero values mainly
on the diagonal $r=q$, top inset), while the lower band consists
of states in which the two particles have small probability to occupy
the same site (lower inset). The gap is proportional to the interaction
strength $U$. Attractive interactions $U<0$, yield an identical
yet inverted spectrum.}

\end{figure}

\emph{The two particle spectrum.-} To understand these results we
consider the two-particle spectrum of the system, as shown in Fig.~\ref{fig: 2}
for two particles on a lattice with $M=29$ sites. Each of the two-particle
eigenfunctions can be written as $\Psi(r_{1},r_{2})$ where $r_{1}$
and $r_{2}$ are the positions of the two particles on the lattice.
Introducing the center of mass coordinate, $R=(r_{1}+r_{2})/2$ and
the relative coordinate, $r=r_{1}-r_{2}$, we can solve the Schrödinger
equation with the ansatz $\Psi(r_{1},r_{2})=\exp(iKR)\psi_{K}(r)$,
where $K$ is the quasi-momentum of the center of mass motion and
$\psi_{K}(r)$ is the pair wavefunction \cite{Maldague}.

For finite interaction strength \textbf{$U$ }the spectrum separates
into two bands. The main part of the spectrum, containing $[M(M-1)]/2$
eigenstates, consists of scattering states having low probability
at $r=0$, whose energy is given by the non-interacting part of the
Hamiltonian. The smaller part of the spectrum ($M$ eigenstates) consists
of states $\psi_{K}^{{\rm bs}}(r)$ which have a large probability
for two particles to occupy the same site, i.e., $|\psi_{K}^{{\rm bs}}(0)|^{2}\rightarrow1\;(U\rightarrow\infty)$
\cite{Maldague} (see insets in Fig.~\ref{fig: 2}). This mini-band
has higher or lower energies than the main part of the spectrum, depending
on the sign of the interaction; nevertheless, the spatial probability
distribution of the two-particle eigenstates is identical.

Using this picture, it is possible to explain the results in Fig.~\ref{fig: 1}.
An initial state in which the two particles occupy the same site with
strong attractive or repulsive interaction, will mostly contain two-particle
states from the smaller mini-band. As a result, the two particles
will remain bound as described by Winkler et al. in \cite{Winkler}
(see Fig.~\ref{fig: 1} g-l). A complementary process happens if
the particles initially occupy different sites. This initial condition
excites mainly scattering states from the main part of the spectrum.
As a result, the particles have low probability to be found at the
same site throughout the evolution, and will not show bunching.

Let us now turn to the case of strong interactions $|U|\gg J$. Our
goal is to understand the {}``fermionization'' as observed in the
correlator $\Gamma_{q,r}(t)$ for an initial state in which the particles
are found at different sites. We start by noting that by focusing
on the scattering states we can describe the Hamiltonian (\ref{eq:CH})
using hard-core bosons, where doubly occupied sites are eliminated
from the Hilbert space. Formally we replace the bosonic with spin-1/2
operators: $a_{m}^{\dag}\rightarrow S_{m}^{+},$ $a_{m}\rightarrow S_{m}^{-}$
.

Next, we use a standard mapping from spin-1/2 to fermionic operators
$f_{m}$, $f_{m}^{\dag}$ \cite{Jordan}. Let us review the essential
steps of this mapping to understand the {}``fermionic'' behavior
of $\Gamma_{q,r}(t)$. Spin-1/2 and fermionic operators share the
local property $(f_{m}^{\dag})^{2}=f_{m}^{2}=(S_{m}^{+})^{2}=(S_{m}^{-})^{2}=0$.
However, spins on different sites commute, whereas fermions pick up
a minus sign. In the sought mapping one corrects for this via the
Jordan-Wigner string $\exp(i\phi_{m})$: \begin{align*}
S_{m}^{-} & =e^{-i\phi_{m}}f_{m},\quad S_{m}^{+}=e^{i\phi_{m}}f_{m}^{\dag},\\
\mbox{with}\qquad\phi_{m} & =\pi\sum_{l=1}^{m-1}f_{l}^{\dag}f_{l}\end{align*}
 It is now straight-forward to check that for $\Gamma_{q,r}(t)$ the
Jordan-Wigner string drops out. Hence, the correlation for hard-core
bosons are identical to the ones obtained for non-interacting fermions
in accordance with our observation in Fig.~\ref{fig: 1}.

\emph{Experimental results.-}\textbf{ }The case of the two-photon
quantum walk and the resulting HBT correlations were considered in
\cite{Bromberg} and observed in \cite{Peruzzo}, in a system of waveguide
lattices. This system is described by an equation identical to Eq.
1, only that for single photons interactions are negligible, i.e.
$U=0$. Ref. \cite{Bromberg} presented also a measurement of correlation
for classical, thermal light waves, analogous to the intensity correlations
predicted by the original HBT work \cite{HBT}. That experiment showed
that the results for thermal inputs captures some (but not all) aspects
of the correlations predicted for the quantum, two-particle case.\textcolor{blue}{ }

The waveguide lattice used in that experiment, described in detail
in \cite{Bromberg,hagai} , has been shown to support nonlinear effects
for high intensity classical light \cite{WGA}. In this case the system
is described by the Hamiltonian:

\begin{equation}
H=-J\sum_{\langle l,m\rangle}\Psi_{l}^{*}\Psi_{m}+\gamma\sum_{m}|\Psi_{m}|^{4}\label{eq:CH}\end{equation}
which is the classical or mean-field limit of Eq. 1. It is therefore
reasonable to presume that the classical correlations for nonlinear
thermal waves in this system will correspond to the results for two
interacting quantum particles. Indeed, In the experiments described
below we find that in the limit of weak interactions the measured
classical correlation for nonlinear thermal waves are similar to the
predicted quantum correlations.

\begin{figure}
\includegraphics[clip]{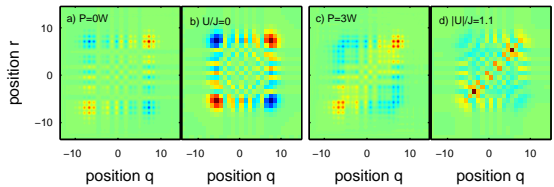}

\caption{\label{fig: 2-1}(color online). Experimental measurements of the
fluctuation in intensity correlations for nonlinear thermal light.
(a) The fluctuations in the intensity correlations in the linear case,
when two thermal beams are injected into two adjacent sites, corresponding
to $|\psi_{initial}\rangle=a_{1}^{\dagger}a_{0}^{\dagger}|0\rangle$.
The results show spatial bunching \cite{Bromberg}. (b) the predictions
of the quantum theory for two, non-interacting particles initially
placed at the same locations. (c) Experimental results for nonlinear
thermal waves (d) the predictions of the quantum theory for two interacting
particles. Note the similarity between the classical results and the
quantum prediction in both cases.}

\end{figure}

\begin{figure}
\includegraphics[clip]{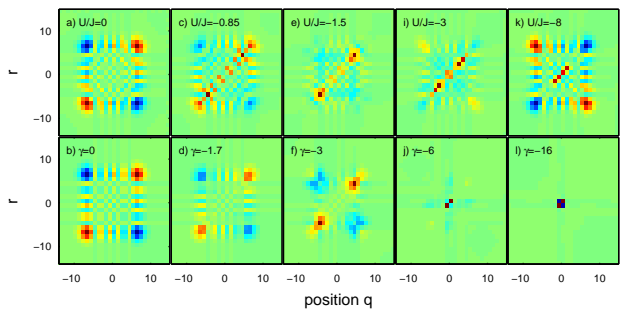}

\caption{\label{fig: 2-2}(color online). Simulations comparing the fluctuations
for the quantum (top panel) and the classical (bottom) cases, with
increasing $|U$| or nonlinear coefficient $|\gamma|$, correspondingly,
for $|\psi_{initial}\rangle=a_{1}^{\dagger}a_{0}^{\dagger}|0\rangle$.
The two sets of results look similar for weak interactions (a-c).
However, beyond $|U|=1.5$ (d-g) the two results diverge - the quantum
correlations transform to fermionic-like correlations, while the classical
correlations become increasingly localized. }

\end{figure}

In Fig. \ref{fig: 2-1} we present experimental measurements for intensity
correlations obtained using nonlinear thermal waves in $|\psi_{initial}\rangle=a_{1}^{\dagger}a_{0}^{\dagger}|0\rangle$.
Detailed numerical results are presented in Fig \ref{fig: 2-2}. In
all figures we compare the correlation fluctuations: $\Gamma_{q,r}^{F}(t)=\left\langle a_{q}^{\dagger}a_{r}^{\dagger}a_{r}a_{q}\right\rangle -\frac{1}{2}\left\langle a_{q}^{\dagger}a_{q}\right\rangle \cdot\left\langle a_{r}^{\dagger}a_{r}\right\rangle =\Gamma_{q,r}-\frac{1}{2}n_{q}\cdot n_{r}$
which are a better basis for comparison between the quantum and classical
(thermal) case. As the results show, for weak interactions the classical
HBT correlations follow the quantum predictions - see Fig. \ref{fig: 2-1}
and Fig. \ref{fig: 2-2} a-f. However, as interactions become stronger
the two systems diverge - while the quantum system exhibits a switch
to fermion-like correlations, the classical system cannot follow,
and remains with modified, localized correlations - Fig. \ref{fig: 2-2}
i-l.

\emph{Proposed cold atoms experiment.-} Experimentally, the strong
interaction limit of effects discussed above can be observed using
techniques that recently became available. The experimental
requirements include the ability to: (i) initially localize exactly
two quantum particles at two pre-determined lattice sites, (ii) to
allow these particles to freely tunnel and exhibit a QW, (iii) to
control the interaction strength (or the interaction-to-tunneling
ratio), and (iv) to image single particles in the lattice sites
after some evolution time. An example of a system that can
accommodate such experiments was recently presented in \cite{Bloch}.
In this system, the authors placed single atoms at selected sites
and allowed them to tunnel, exhibiting, on the ensemble average, the
dynamics of continuous time QWs \cite{Qwalk,hagai}. Using a similar
approach, we propose starting with an ensemble of two (and in
principle N) atoms separated by several sites. The density after
time T can be measured in the same manner as in \cite{Bloch}, and
the two (or N) particle correlation can be directly assessed from
the raw data. An important aspect will be the ability to control
interactions, for example via tuning a Feshbach resonance, or
controlling the ratio $U/J$. For the two particle case and at zero
interactions, the results should correspond to those presented in
\cite{Bromberg} and observed in \cite{Peruzzo} using photon pairs.
However, when interactions will be introduced we predict the results
presented above: if the particles are places one on top of each
other they will tunnel as a pair \cite{Winkler}, and the density
will become localized. If they are placed at different, not too
distant sites, the density will show only minor changes as a
function of $U$, but the two-particle correlation will change
significantly, reaching a fermion-like correlations at the limit of
strong interactions. In the same spirit, this system can be used to
directly measure the dynamic properties and correlations for large
number of particles, a problem which quickly becomes impossible to
compute.

\emph{Conclusions}\textbf{\emph{.-}} In this letter we have
considered the quantum dynamics of two bosons on a lattice, each
initially confined to a single site. Such dynamics with two or more
particles can be experimentally explored in systems such as
described in ref. \cite{Bloch}. As the number of particles increases
the problem will become uncomputable, but may remain experimentally
accessible.

\emph{Note.- }During the final completion of this manuscript we became
aware of related work \cite{related}.


\begin{thebibliography}{20}
\bibitem{Qwalk}J. Kempe, Contemp. Phys., \textbf{44}, 307, (2003).

\bibitem{Photosynthesis}G. S. Engel, et al., Nature \textbf{446},
782-786 (2007).

\bibitem{Takuya}T. Kitagawa, et al., Phys. Rev. A \textbf{82}, 033429
(2010).

\bibitem{UniversalQC}A. M. Childs, Phys. Rev. Lett. \textbf{102},
180501 (2009).

\bibitem{hagai}H. B. Perets, et al., Phys. Rev. Lett. \textbf{100},
170506 (2008).

\bibitem{Silberhorn}A. Schreiber et al., Phys. Rev. Lett. \textbf{104},
050502 (2010).

\bibitem{White}M. A. Broome, et al., Phys. Rev. Lett. \textbf{104},
153602 (2010).

\bibitem{atomQwalk} M. Karski, et al., Science \textbf{325}, \textbf{174}
(2009).

\bibitem{Bloch}C. Weitenberg, et al., arxiv:1101.2076 (2011).

\bibitem{Bromberg}Y. Bromberg et al., Phys. Rev. Lett. \textbf{102},
253904 (2009).

\bibitem{Peruzzo} A. Peruzzo, et al., Science \textbf{329}, 1500
(2010).

\bibitem{HBT} R. Hanbury Brown and R.Q. Twiss, Nature \textbf{177},27
(1956).

\bibitem{Havlin}This problem can be considered as the quantum counterpart
of classical excluded volume diffusion which is non-trivial already
in 1D lattices due to the inability to neglect fluctuations; see e.g.
D. ben-Avraham and S. Havlin, Diffusion and Reactions in Fractals
and Disordered Systems, chapter 10. (Cambridge University Press, 2000).

\bibitem{Winkler}K. Winkler, et al., Nature \textbf{441}, 853 (2006).

\bibitem{Maldague}P. F. Maldague, Phys. Rev. B \textbf{16}, 2436
(1977).

\bibitem{Sup} See xxxx for supplementary information.

\bibitem{Jordan} P. Jordan and E. Wigner, Z. Phys.\textbf{ 47}, 631
(1928)

\bibitem{WGA}D. N. Christodoulides, F. Lederer, and Y. Silberberg,
Nature \textbf{424} 424, 817 (2003); F. Lederer, et al., Physics Reports
\textbf{463}, 1 (2008).

\bibitem{NLHBT} Y. Bromberg, Y. Lahini, E. Small, and Y. Silberberg,
Nat. Photon. \textbf{4}, 721(2010).

\bibitem{related}A. Ahlbrecht, et al., arXiv:1105.1051 (2011).
\end{thebibliography}
\end{document}